\documentclass[cameraready]{Interspeech}
\usepackage{cite}
\usepackage{subcaption}

\title{Causal Tracing of Audio-Text Fusion in Large Audio Language Models}

\author[affiliation={1}]{Wei-Chih}{Chen}
\author[affiliation={2}]{Chien-yu}{Huang}
\author[affiliation={1,3}]{Hung-yi}{Lee}

\address{
    $^1$ National Taiwan University, Taiwan \\
    $^2$ Carnegie Mellon University, United States
    \\
    $^3$ NTU Artificial Intelligence Center of Research Excellence
}

\email{timmy90617055@gmail.com}


\usepackage{comment}

\begin{document}

\maketitle

%
\begin{abstract}
{
Despite the strong performance of large audio language models (LALMs) in various tasks, exactly how and where they integrate acoustic features with textual context remains unclear.
We adapt causal tracing to investigate the internal information flow of LALMs during audio comprehension.
By conducting layer-wise and token-wise analyses across DeSTA, Qwen, and Voxtral, we evaluate the causal effects of individual hidden states.
Layer-wise analysis identifies different fusion strategies, from progressive integration in DeSTA to an abrupt late-stage fusion in Qwen.
Token-wise analysis shows that the final sequence token acts as an informational bottleneck where the network decisively retrieves relevant information from the audio.
We also observe an attention-like query mechanism at intermediate token positions that triggers the model to pull task-relevant audio context.
These findings provide a clear characterization of when and where multi-modal integration occurs within LALMs.
}

\end{abstract}
\section{Introduction}

%
{
Recent progress in large language models (LLMs)~\cite{grattafiori2024llama, bai2023qwen, qwen2, qwen2.5, qwen3, liu2026ministral} has enabled multi-modal systems to perceive and reason over various sensory inputs.
Among these, Large Audio Language Models (LALMs)~\cite{lu2025developing, lu2025desta2, chu2023qwen, chu2024qwen2, liu2025voxtral, gong2023joint, tang2023salmonn, hu2024wavllm} now achieve strong performance on tasks ranging from speech recognition and captioning~\cite{guo2025brace, yang2024air, huang2025dynamicsuperb} to complex audio-aware question answering~\cite{ma2025mmar, yang2025sakura}.
These models typically integrate audio encoders~\cite{radford2023robust, chen2022beats} with LLM backbones~\cite{grattafiori2024llama, bai2023qwen, qwen2, qwen2.5, qwen3, liu2026ministral} to jointly process continuous audio streams and text inputs, mapping auditory features directly into the linguistic embedding space of the LLM.
}


{
Despite this empirical success, the internal mechanisms driving this multi-modal integration remain unclear.
Current research primarily focuses on architectural design and macroscopic task-level evaluations, such as overall performance~\cite{guo2025brace, huang2024dynamic, huang2025dynamicsuperb, yang2021superb, yang2025sakura, yang2024air, sakshi2024mmau, ma2025mmar}, bias~\cite{lin2024listen, lin2024spoken}, and hallucination~\cite{kuan2025can, cheng2025ahabench}.
As a result, the exact dynamics of audio-text fusion are still a black-box.
We currently lack a precise understanding of when across the model's depth this integration occurs, such as early versus late-stage fusion, and where across the input sequence the network causally bridges auditory inputs with textual reasoning.
}


{
Uncovering these dynamics is essential for improving model interpretability~\cite{meng2022locating}, mitigating hallucinations~\cite{agrawal2024language}, and guiding architectural design~\cite{michel2019sixteen}.
To investigate how these models function internally, mechanistic interpretability provides interventional techniques such as causal tracing to map hidden information flow~\cite{meng2022locating}.
While these methods have successfully traced reasoning pathways in text models~\cite{meng2022locating, zhang2023towards} and visual alignments in vision-language systems~\cite{palit2023towards, li2025causal}, their application to large audio language models remains limited.
Extending these tools to track how auditory attributes are integrated with textual context is a critical next step for understanding multi-modal reasoning.
}


{
This paper investigates these internal dynamics by systematically adapting \textit{causal tracing} to track how acoustic features are incorporated into linguistic representations.
We propose a framework utilizing both spatial and depthwise analyses to evaluate the causal effects of specific hidden states during audio comprehension. By tracing the flow of information across different layers and token positions, we characterize the distinct pathways the network employs to bridge input audio with its textual output.

We conduct experiments across several model families, including DeSTA, Qwen, and Voxtral.
The results show that: (1) we identify divergent fusion strategies, ranging from progressive integration in DeSTA architectures to polarized late-stage fusion in Qwen models (Sec. \ref{sec:fusion_depth}); (2) the final sequence token acts as a critical informational bottleneck where the network retrieves relevant information from the audio to generate its response (Sec. \ref{sec:fusion_spatial}); (3) an attention-like query mechanism exists at intermediate token positions that triggers the model to extract specific acoustic features (Sec. \ref{sec:fusion_spatial}); and (4) the architectural divergences observed offer actionable insights for future model design and deployment (Sec. \ref{sec:implication}).

\begin{figure}[t]
    \centering
    \includegraphics[width=\columnwidth]{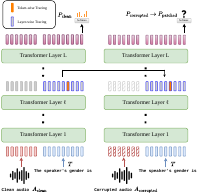}
    \caption{
    Illustration of our tracing methodology. The clean run (left) processes the original audio, while the corrupted run (right) replaces the audio with silence. By patching specific hidden states from the clean run into the corrupted run (e.g., layer-wise or token-wise), we isolate the causal contribution of individual layers and sequence positions to the final prediction.
    }
    \label{fig:overview}
\end{figure}
\section{Related Works}

{
\subsection{Probing Speech and Audio Models} 
Several studies \cite{raj2019probing, belinkov2017analyzing, pasad2021layer, pasad2023comparative, choi2024self, lin2024property, wu2024and} have investigated the internal mechanics of speech and audio models, ranging from early architectures \cite{xvectors, amodei2016deep} to self-supervised learning frameworks \cite{hsu2021hubert, chang2022distilhubert, schneider2019wav2vec, baevski2020wav2vec, conneau2020unsupervised}.
Detailed layer-wise \cite{pasad2021layer, pasad2023comparative, choi2024self} and neuron-wise analyses \cite{lin2024property, wu2024and} have been widely used to understand how these models process acoustic features.
While the internal dynamics of these unimodal models are well documented, modern investigations into LALMs \cite{guo2025brace, huang2024dynamic, huang2025dynamicsuperb, yang2021superb, yang2025sakura, yang2024air, sakshi2024mmau, ma2025mmar} rarely examine their internal working mechanisms.
Instead of analyzing the cross-modal fusion process, current research predominantly focuses on task-level performances.
Recent benchmarks primarily assess overall performance \cite{guo2025brace, huang2024dynamic, huang2025dynamicsuperb, yang2021superb, yang2025sakura, yang2024air, sakshi2024mmau, ma2025mmar}, model bias \cite{lin2024listen, lin2024spoken}, or hallucination \cite{kuan2025can} based solely on the final generated text.
Since these evaluations treat the LALM as an end-to-end black-box, the fundamental question of how and where acoustic features integrate with text representations remains unanswered.
}


%
{
\subsection{Mechanistic Interpretability and Causal Tracing}
Mechanistic interpretability investigates how neural networks produce predictions by analyzing their internal hidden states.
These analyses generally fall into two methodological categories: observational and interventional.
Observational methods evaluate hidden states without altering the model computation, revealing only correlations between representations and specific tasks. In contrast, interventional methods, such as causal tracing~\cite{zhang2023towards}, actively modify hidden states during the forward pass to establish direct causal relationships.

This causal approach has been highly successful in natural language processing.
For example, the ROME framework~\cite{meng2022locating} utilizes hidden state interventions to isolate the exact feed-forward layers causally responsible for factual recall.
These interventional tools have recently been extended to multimodal systems~\cite{palit2023towards, li2025causal, golovanevsky2024vlms, chien2026mint}.
For vision language models, causal techniques have identified specific layers where visual representations integrate with textual context~\cite{chien2026mint}.

However, interpretability in the speech and audio domain still heavily relies on observational methods.
For instance, AudioLens~\cite{yang2025audiolens} employs the logitlens technique, which projects intermediate hidden states directly to the vocabulary space, to observe layer-wise representation shifts.
While these observational methods provide valuable correlational insights, they cannot isolate the exact hidden states that causally drive multimodal integration.
This paper addresses the limitation by adapting causal tracing to actively intervene on LALMs, explicitly evaluating how acoustic features integrate into the linguistic embedding space.
}
\begin{figure*}[t!]
    \centering
    
    \begin{subfigure}[b]{0.24\textwidth}
        \centering
        \includegraphics[width=\textwidth]{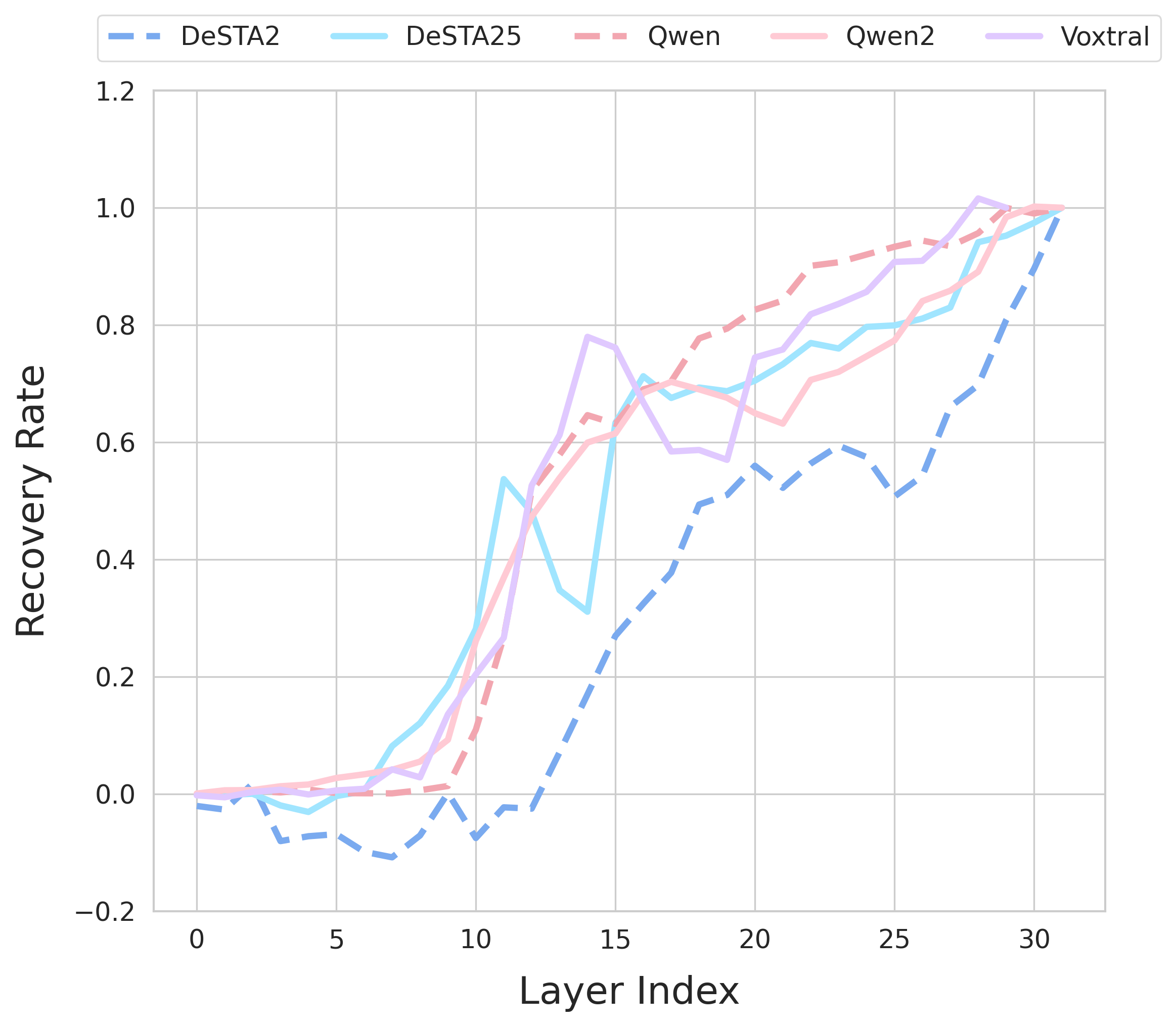}
        \caption{Animal}
        \label{fig:animal_layer}
    \end{subfigure}
    \hfill 
    \begin{subfigure}[b]{0.24\textwidth}
        \centering
        \includegraphics[width=\textwidth]{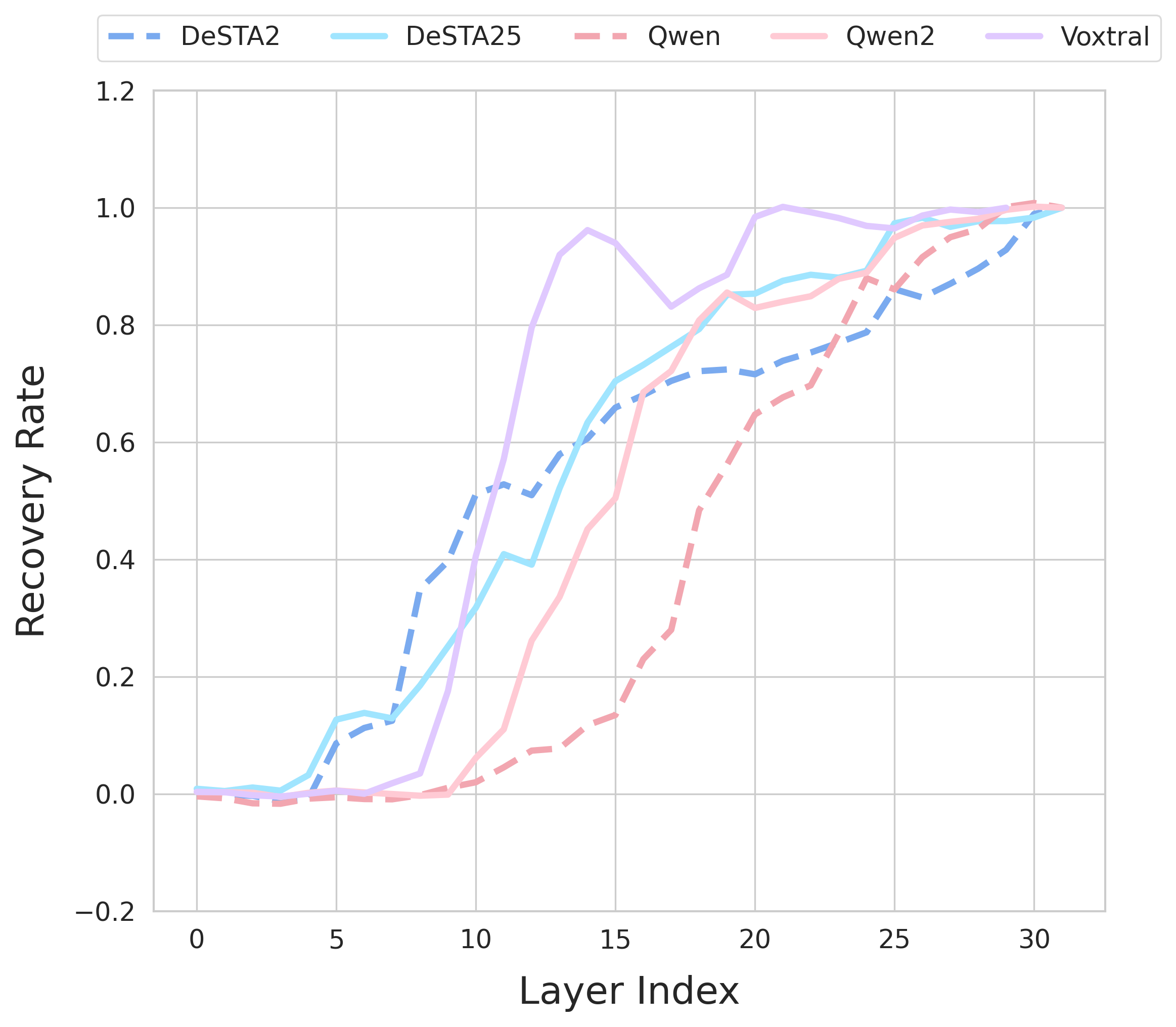}
        \caption{Emotion}
        \label{fig:emotion_layer}
    \end{subfigure}
    \hfill 
    \begin{subfigure}[b]{0.24\textwidth}
        \centering
        \includegraphics[width=\textwidth]{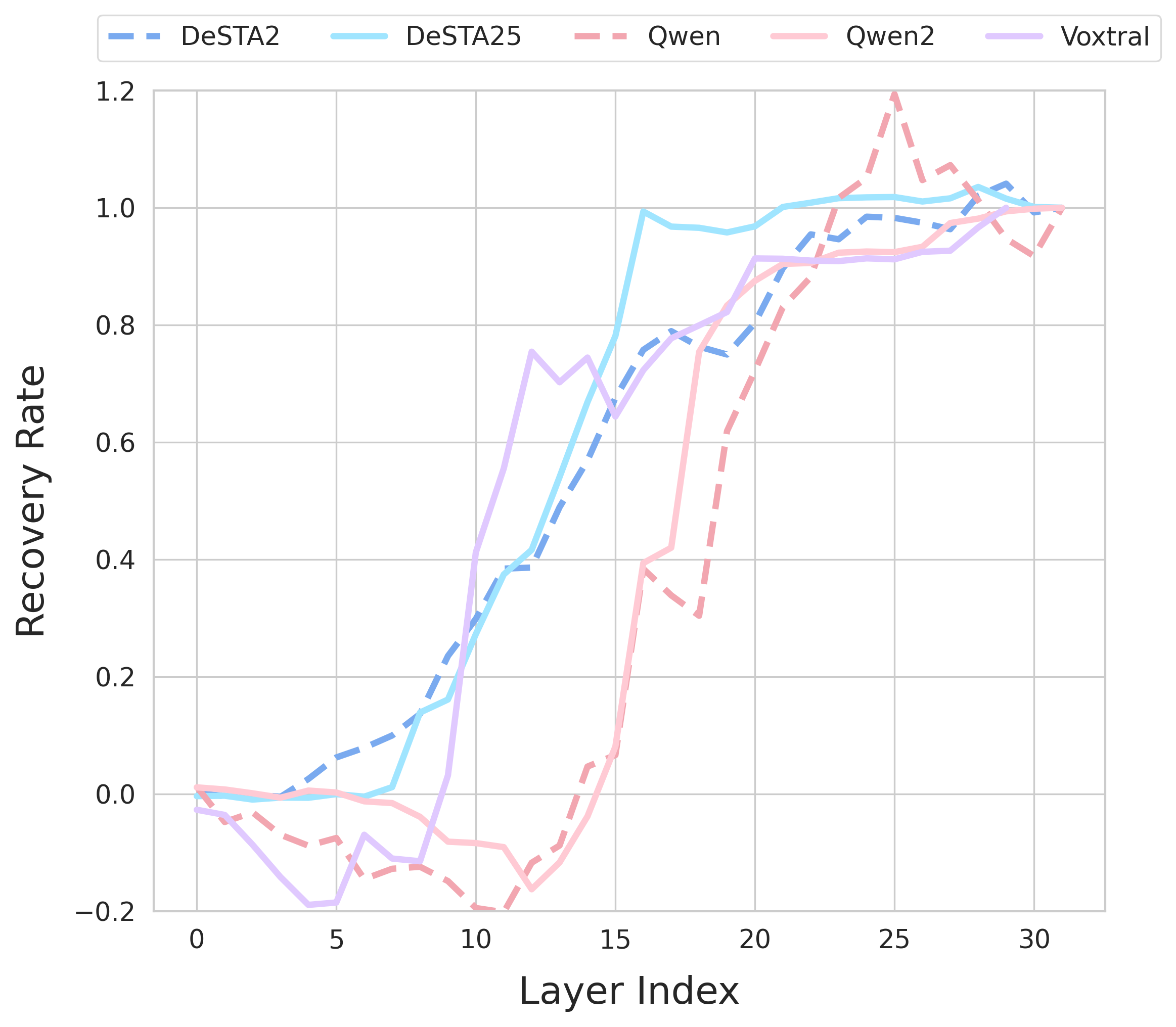}
        \caption{Gender}
        \label{fig:gender_layer}
    \end{subfigure}
    \hfill 
    \begin{subfigure}[b]{0.24\textwidth}
        \centering
        \includegraphics[width=\textwidth]{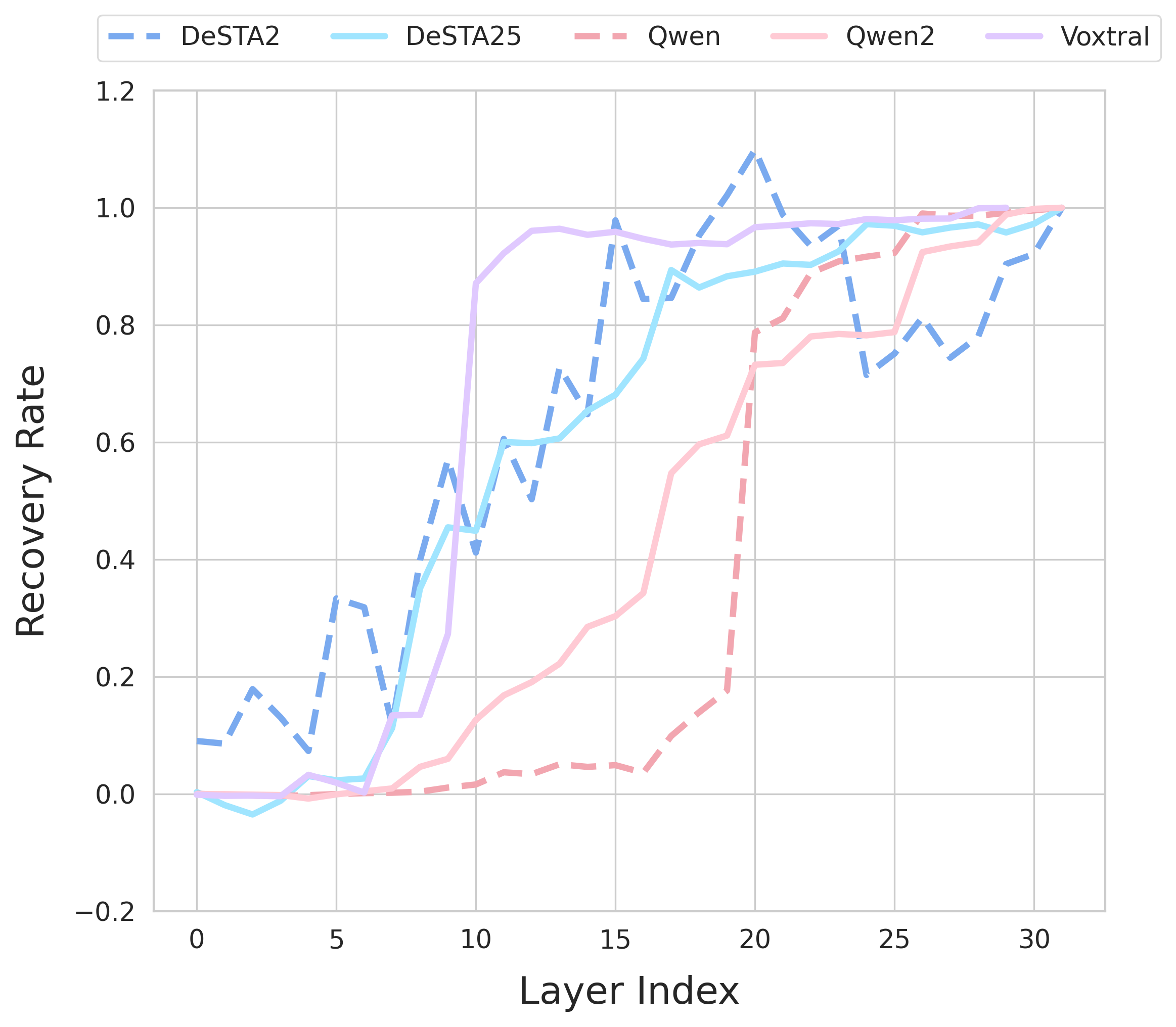}
        \caption{Language}
        \label{fig:language_layer}
    \end{subfigure}
    
    \caption{Layer-wise tracing results for all models across four auditory attributes. The results reveal distinct multi-modal integration strategies across model families. DeSTA architectures show a progressive fusion approach. Qwen models demonstrate an abrupt late-stage fusion pattern. Meanwhile, Voxtral exhibits an early fusion strategy.}
    \label{fig:layer_wise_result}
\end{figure*}

\section{Method}

\subsection{Setup}
{

To isolate the causal effect of the audio input on the text generation, we adapt the causal tracing framework of ROME~\cite{meng2022locating}, formalizing the process using three distinct inference runs, as illustrated in Figure~\ref{fig:overview}. Let $A$ represent the input audio sequence, $T$ the task specific textual prompt, and $y$ the correct target answer token. The model calculates the next token probability distribution, yielding the probability of the target answer $P(y | A, T)$.
\begin{itemize}
    \item \textbf{Clean run:} The model processes the original audio $A_{\text{clean}}$ and prompt $T$. The resulting target probability is defined as $P_{\text{clean}} = P(y | A_{\text{clean}}, T)$.
    \item \textbf{Corrupted run:} The input audio is completely replaced by a baseline of silence, denoted as $A_{\text{corrupt}}$. This removes the acoustic context, yielding the probability $P_{\text{corrupted}} = P(y | A_{\text{corrupt}}, T)$.
    \item \textbf{Patched run:} During the forward pass of a corrupted run, we intervene by replacing a selected set of hidden states $h_{\text{corrupt}}$ with the exact corresponding hidden states $h_{\text{clean}}$ cached from the clean run. The target probability after applying this intervention is denoted as $P_{\text{patched}}$.
\end{itemize}

The selection of pure silence as the corrupted baseline is a critical and highly effective methodological choice for this experimental setup. In mechanistic interpretability, a robust  baseline must cleanly ablate the target information without introducing out-of-distribution artifacts that could induce model hallucinations. Using pure silence achieves exactly this criterion. It acts as a strict and neutral zero-information regarding the evaluated acoustic attributes. By replacing the speech waveform with silence, we completely remove the acoustic information while perfectly preserving the prompt's structural integrity, sequence length, and audio token count. This deliberate isolation ensures that any change in the model's prediction is driven exclusively by the absence of the relevant multi-modal signal, establishing a highly reliable foundation for measuring causal effects.

Following~\cite{li2025causal}, we quantify the causal impact of these patched hidden states using the Recovery Rate (RR) metric:
$$\text{RR} = \frac{P_{\text{patched}} - P_{\text{corrupted}}}{P_{\text{clean}} - P_{\text{corrupted}}}$$

This metric measures the proportion of the target probability restored by the intervention.
An RR value approaching $1.0$ indicates that the patched neural component contains sufficient causal information to recover the original prediction, whereas a value near $0.0$ implies no causal contribution to bridging the modalities.
To ensure metric validity, we exclude samples where the model predicts incorrectly in the clean run, correctly in the corrupted run, or where $P_{\text{clean}} \le P_{\text{corrupted}}$.

To thoroughly investigate multi-modal integration within LALMs, analyzing the network across a single dimension is insufficient.
We evaluate the internal dynamics along two distinct axes: the vertical depth of the transformer layers (layer-wise) and the horizontal sequence of the text tokens (token-wise).
We utilize both layer-wise and token-wise causal tracing to construct a complete causal map.
This dual approach allows us to explicitly answer \textit{when} the semantic integration occurs across the model's depth and \textit{where} the relevant audio context is routed among specific token positions to generate the final output.
}

\subsection{Layer-wise Tracing}


{

Layer-wise tracing patches the hidden states of all textual tokens across a single layer from the clean run into the corresponding layer of the corrupted run. Formally, let $h_{\text{clean}}^{(l)}$ denote the hidden states of all text tokens at the $l$-th transformer layer in the clean run. During the intervention, we replace the corrupted hidden states $h_{\text{corrupt}}^{(l)}$ with $h_{\text{clean}}^{(l)}$.
Crucially, after this substitution, the forward pass continues through all subsequent transformer layers using this patched representation.
In Figure~\ref{fig:overview}, this operation is denoted by the dark blue blocks.

The primary motivation for this technique is to trace the temporal flow of influence across the depth of the network.
By observing how the recovery rate changes layer by layer, we identify the exact depth at which semantic integration occurs.
This helps us determine whether different architectures process modalities independently before a late stage fusion or integrate them progressively.
}

\subsection{Token-wise Tracing}
\label{sec:token-tracing}

{

While layer-wise analysis reveals the depth of integration, it lacks granular spatial resolution. Token-wise tracing isolates and patches a single hidden state at a specific textual token position. Formally, we intervene by replacing $h_{\text{corrupt}, i}^{(l)}$ with $h_{\text{clean}, i}^{(l)}$, which represents the hidden state at the $l$-th layer for the $i$-th token. The network then completes the forward pass using this localized patch.
Figure~\ref{fig:overview} illustrates this intervention by the solid orange block.

Token-wise tracing provides a precise spatial map of where modalities intersect across the textual prompt.
To systematically study the spatial distribution of causal effects, we follow~\cite{li2025causal} in explicitly defining the functional regions of our input sequence.
Using a gender classification prompt (\textit{``What is the gender of the speaker in the speech? Possible options: male, female. The speaker's gender is''}) as an example, we partition the text into four distinct segments:

\begin{enumerate}
    \item \textbf{Early prompt tokens:} The preliminary instructions setting up the task (e.g., \textit{``What is the gender of the speaker in the speech? Possible options:''}).
    \item \textbf{Object tokens:} The specific class labels or target attributes provided as choices (e.g., \textit{``male, female''}).
    \item \textbf{Late prompt tokens:} The transitional text that follows the object tokens and sets up the final answer (e.g., \textit{``The speaker's gender''}).
    \item \textbf{The last token:} The absolute final token of the given input sequence (e.g., \textit{``is''}). Because LALMs generate text autoregressively, the probability distribution for the target answer is computed directly from the hidden state at this exact final position.
\end{enumerate}

By mapping interventions to these specific categories, we investigate whether the model continuously blends signals throughout the sequence or decisively retrieves audio context at specific informational bottlenecks.

}
\section{Experimental Settings}

\subsection{Dataset}


{
We use the SAKURA dataset~\cite{yang2025sakura} to evaluate the model's perception of distinct acoustic attributes.
We focus on four target attributes: (1) animal, (2) emotion, (3) gender, and (4) language.
Following AudioLens~\cite{yang2025audiolens}, we format the input for these classification tasks as a structured question answering prompt. The model receives the input audio paired with a textual prompt that defines the target attribute and the available options (e.g., \textit{What is the gender of the speaker in the speech? Possible options: male, female.}).
To constrain the model's generation, we append a response phrase to the end of the prompt (e.g., \textit{The speaker's gender is}).
This format forces the model to generate the next token probability distribution to complete the sentence.
Consequently, we can accurately trace how the relevant audio context is retrieved and processed to select the correct attribute target.

}

\subsection{Models}

{
We evaluate the DeSTA2 \cite{lu2025developing}, DeSTA2.5 \cite{lu2025desta2}, Qwen \cite{chu2023qwen}, Qwen2 \cite{chu2024qwen2}, and Voxtral~\cite{liu2025voxtral} models.
These models achieve leading performance on recent audio reasoning benchmarks and represent distinct structural families.
They utilize different combinations of audio encoders, cross modal adapters, and LLM backbones to integrate acoustic features into the linguistic embedding space.
Comparing these diverse architectures provides a comprehensive view of internal fusion mechanisms across modern LALMs.
}
\section{Results}

%
    
    
    
    

\begin{figure*}[t!]
    \centering
    
    \begin{subfigure}[b]{0.24\textwidth}
        \centering
        \includegraphics[width=\textwidth]{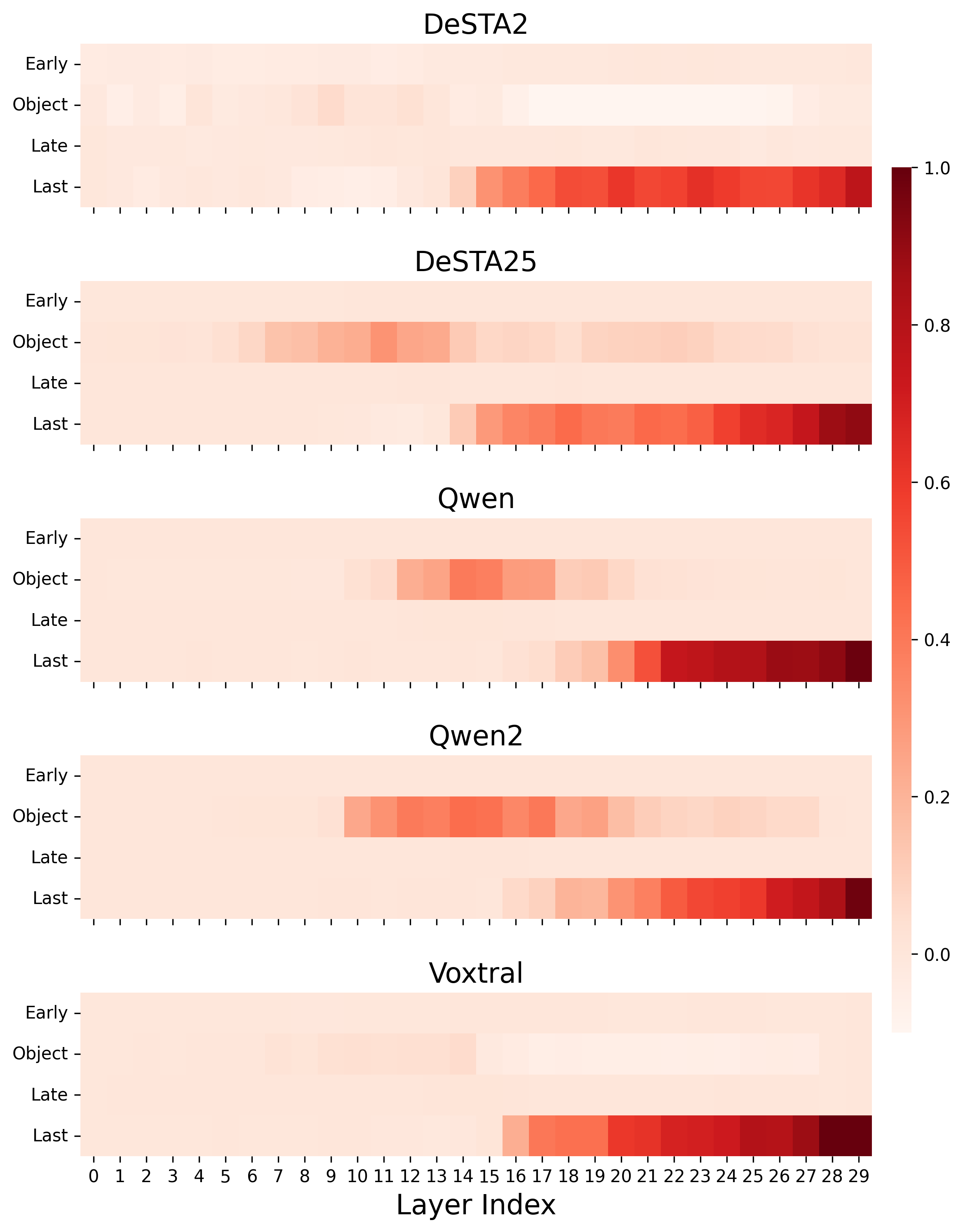}
        \caption{Animal}
        \label{fig:animal_token}
    \end{subfigure}
    \hfill 
    \begin{subfigure}[b]{0.24\textwidth}
        \centering
        \includegraphics[width=\textwidth]{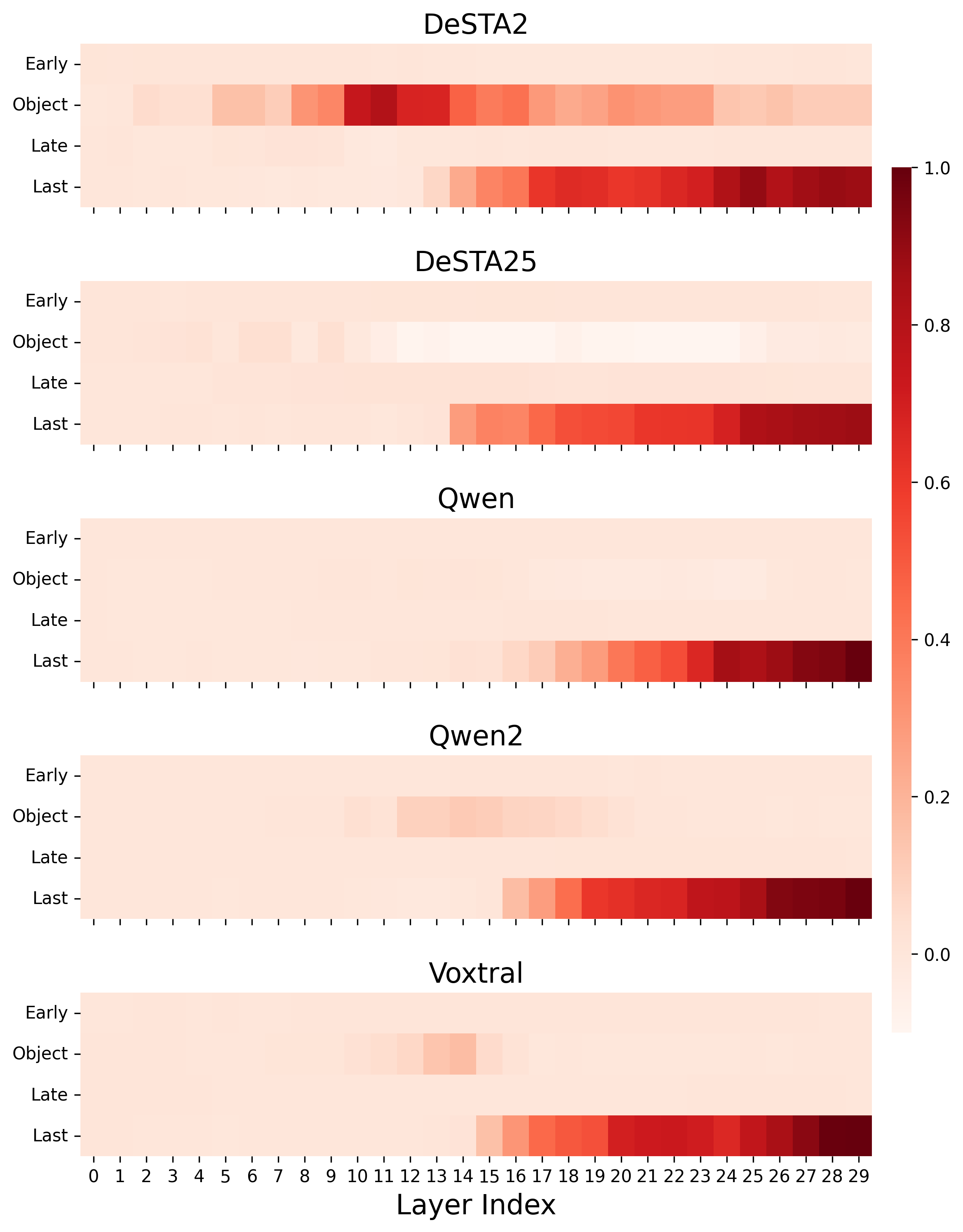}
        \caption{Emotion}
        \label{fig:emotion_token}
    \end{subfigure}
    \hfill 
    \begin{subfigure}[b]{0.24\textwidth}
        \centering
        \includegraphics[width=\textwidth]{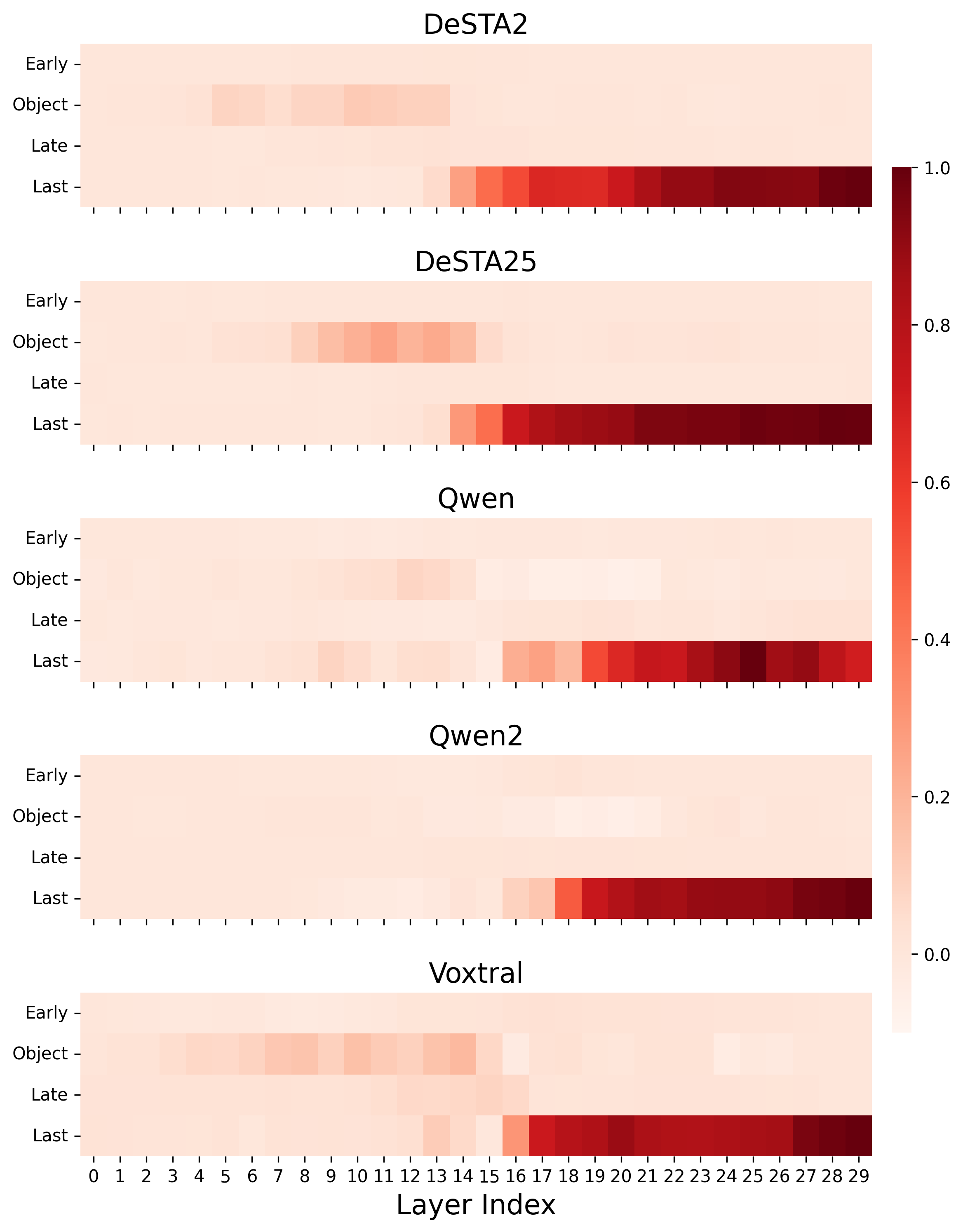}
        \caption{Gender}
        \label{fig:gender_token}
    \end{subfigure}
    \hfill 
    \begin{subfigure}[b]{0.24\textwidth}
        \centering
        \includegraphics[width=\textwidth]{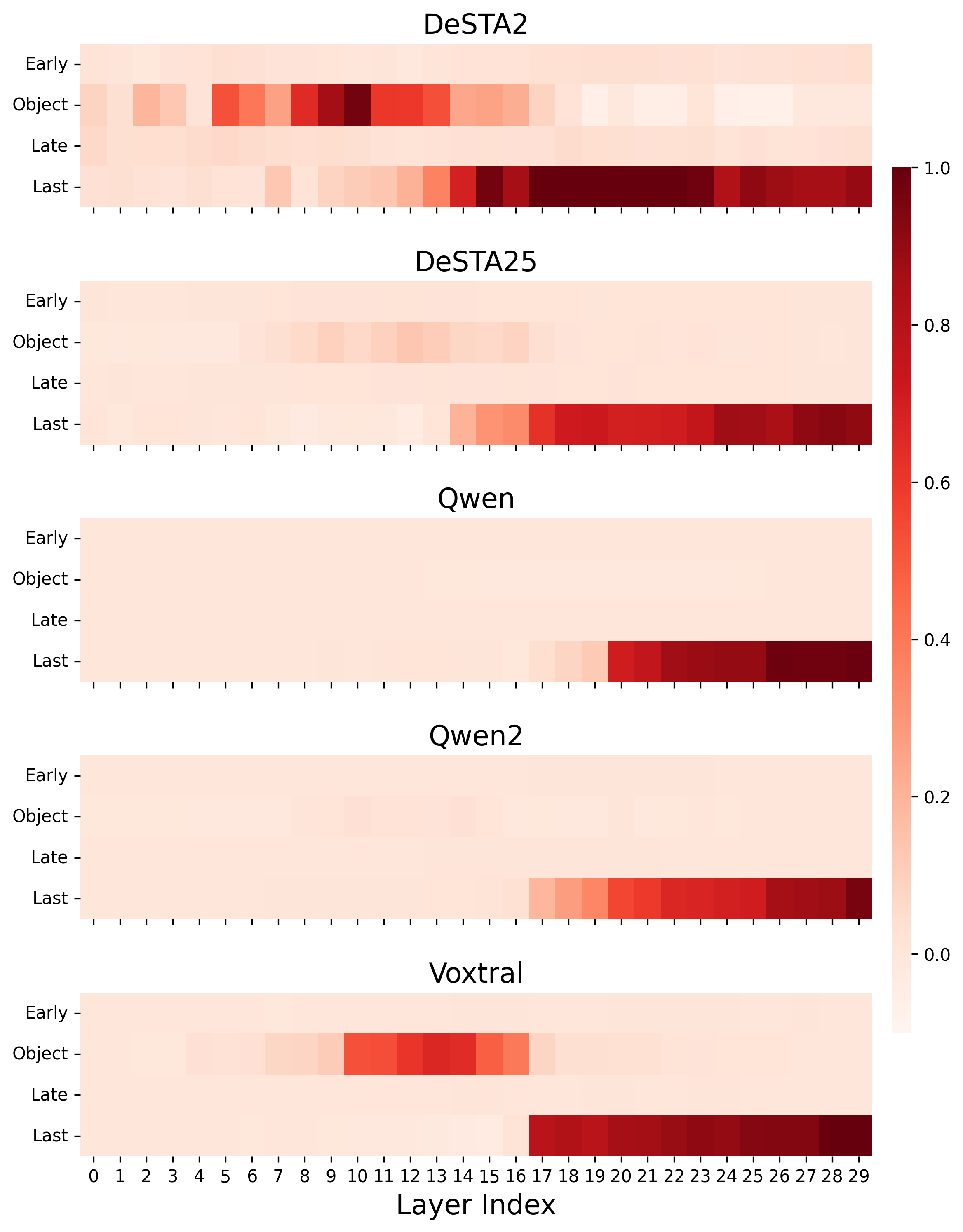}
        \caption{Language}
        \label{fig:language_token}
    \end{subfigure}
    
    \caption{Token-wise tracing results across all models and four auditory attributes reveal the last token as a critical informational bottleneck for audio context retrieval, while object tokens trigger an attention-like query mechanism to extract specific target attributes.}
    \label{fig:token_wise_result}
\end{figure*}

{
We present our causal tracing findings along two dimensions: (1) layer-wise tracing to establish the depth of multi-modal integration (Sec.~\ref{sec:fusion_depth}), and (2) token-wise tracing to identify its spatial localization (Sec.~\ref{sec:fusion_spatial}).
Building on these observations, we discuss the practical implications of these architectural dynamics for future model design (Sec.~\ref{sec:implication}).
}

\subsection{Depth of Multi-Modal Integration}
\label{sec:fusion_depth}
Figure~\ref{fig:layer_wise_result} plots the recovery rate against the layer index, revealing distinct integration strategies among the evaluated models. The DeSTA architectures exhibit a progressive approach to multi-modal fusion. For DeSTA2 and DeSTA2.5, the recovery rate demonstrates a gradual increase as depth progresses, with causal effects stabilizing after layer 15 across most target attributes.

Conversely, the Qwen family (Qwen and Qwen2) exhibits a late stage fusion pattern.
In these models, the causal impact of the audio signal remains near zero throughout the early layers.
A sharp transition to a high recovery rate only occurs within the final third of the network (around layers 18 to 31).
This indicates that Qwen models process textual and acoustic modalities independently for the majority of their depth, deferring semantic integration until the deepest layers of the LLM backbone.

However, Voxtral demonstrates an accelerated recovery trajectory.
Across attributes such as Emotion and Gender, Voxtral achieves maximum recovery significantly earlier in the network compared to both the DeSTA and Qwen families, indicating an early fusion strategy.

\subsection{Spatial Localization of Audio-Text Fusion}
\label{sec:fusion_spatial}
Figure~\ref{fig:token_wise_result} presents the token-wise tracing results, providing a spatial map of exactly where multi-modal fusion takes place across the entire input sequence.
By categorizing the input into early prompt tokens, object tokens, late prompt tokens, and the last token (Sec.~\ref{sec:token-tracing}), we identify two primary mechanisms for retrieving audio context.

Across all evaluated models, the highest recovery rate consistently localizes to the final token immediately preceding generation. This indicates that the last position acts as a critical information bottleneck. Instead of continuously blending signals throughout the sequence, the network retrieves the required audio context at the very end to predict the target answer. Notably, an analogous final-token bottleneck has been reported in vision-language models~\cite{li2025causal}. Its recurrence across the audio and visual modalities suggests a transformer-general retrieval mechanism rather than a modality-specific effect.

In addition to this final token bottleneck, we observe a secondary causal activation at the object token within intermediate layers. For the Qwen models, the Animal attribute results show a localized causal spike at the object token during the middle layers. Similarly, the DeSTA and Voxtral models demonstrate significant causal activity at the object token for Language attributes. This localized activation suggests an attention-like query mechanism, where explicitly processing the specific target attribute triggers the network to pull relevant audio context from the acoustic representations.

\subsection{Implications for LALM Design}
\label{sec:implication}
The architectural divergences observed in our layer-wise analysis provide actionable insights for future model design and deployment. For instance, models exhibiting a late-stage fusion pattern present significant opportunities for computational optimization; by treating the early layers strictly as unimodal text processors, developers can have the potential to reduce computational overhead and inference costs. On the other hand, architectures that employ early or progressive fusion strategies may be more advantageous for tasks requiring fine-grained, continuous interaction between acoustic and textual features.

Beyond depth dynamics, identifying spatial bottlenecks has direct implications for mitigating model hallucinations. Our token-wise tracing indicates that the object token acts as a critical trigger for retrieving audio context. If this causal activation fails to trigger, the model is highly susceptible to guessing based on textual priors rather than the input audio. Monitoring this internal query mechanism could serve as a valuable indicator of hallucination at inference. These spatial insights can guide more effective prompt engineering by explicitly highlighting  where multi-modal models fetch contextual information.
\section{Conclusion}

While Large Audio Language Models achieve strong performance, current evaluations primarily focus on final outputs, leaving how they internally integrate acoustic features with textual context unclear.
This paper introduces a causal tracing framework to explicitly investigate these internal fusion mechanisms.
Through layer-wise and token-wise analyses, we identify distinct architectural strategies for multi-modal integration and map the specific sequence positions where models retrieve relevant audio context.
These findings establish a foundational understanding that may help guide future efforts to optimize model efficiency and enhance architectural transparency.

\section{Generative AI Usage Disclosure}

The authors used generative AI primarily to refine the linguistic clarity of this manuscript.
While AI tools provided minor assistance in writing code, all implementations were manually tested and verified.
The authors take full responsibility for the accuracy, methodology, and overall content of this paper.

\section{Acknowledgement}
This work was supported by the Ministry of Education (MOE) of Taiwan under the project Taiwan Centers of Excellence in Artificial Intelligence, through the NTU Artificial Intelligence Center of Research Excellence (NTU AI-CoRE).

\bibliographystyle{IEEEtran}
\bibliography{mybib}

\end{document}